\documentclass[5p,times]{elsarticle}

\usepackage{amssymb}
\usepackage{amsmath}            
\usepackage{epstopdf}           
\usepackage{flushend}           

\usepackage{cite}
\usepackage{amsmath,amssymb,amsfonts}
\usepackage{amssymb}
\usepackage{algorithmic}
\usepackage{graphicx}
\usepackage{textcomp}
\usepackage{xcolor}
\usepackage{comment}
\usepackage{caption}
\usepackage[normalem]{ulem}
\usepackage{algorithm}
\usepackage{algorithmic}
\usepackage{mathtools}

\usepackage{enumitem}

\begin{document}

\begin{frontmatter}



\title{PrivChain: Provenance and Privacy Preservation in \\Blockchain enabled Supply Chains}


\author[First]{Sidra~Malik\corref{cor1}\fnref{label2}}
\ead{sidra.malik@unsw.edu.au}

\author[Second]{Volkan~Dedeoglu}
\ead{volkan.dedeoglu@data61.csiro.au}

\author[First]{Salil~S.~Kanhere}
\ead{salil.kanhere@unsw.edu.au}

\author[Fourth]{Raja~Jurdak}
\ead{r.jurdak@qut.edu.au}

\address[First]{University of New South Wales, Sydney, NSW, 2052 Australia }
\address[Second]{CSIRO Data61, Pullenvale}
\address[Fourth]{Queensland University of Technology, Brisbane, QLD, 4000 Australia}

\begin{abstract}
Blockchain offers traceability and transparency to supply chain event data and hence can help overcome many challenges in supply chain management such as: data integrity, provenance and traceability.
However, data privacy concerns such as the protection of trade secrets have hindered adoption of blockchain technology. Although consortium blockchains only allow authorised supply chain entities to read/write to the ledger, privacy preservation of trade secrets cannot be ascertained. In this work, we propose a privacy-preservation framework, PrivChain, to protect sensitive data on blockchain using zero knowledge proofs. PrivChain provides provenance and traceability without revealing any sensitive information to end-consumers or supply chain entities. Its novelty stems from: a) its ability to allow data owners to protect trade related information and instead provide proofs on the data, and b) an integrated incentive mechanism for entities providing valid proofs over provenance data. In particular, PrivChain uses Zero Knowledge Range Proofs (ZKRPs), an efficient variant of ZKPs, to provide origin information without disclosing the exact location of a supply chain product. Furthermore, the framework allows to compute proofs and commitments off-line, decoupling the computational overhead from blockchain. The proof verification process and incentive payment initiation are automated using blockchain transactions, smart contracts, and events. A proof of concept implementation on Hyperledger Fabric reveals a minimal overhead of using PrivChain for blockchain enabled supply chains.
\end{abstract}

\end{frontmatter}


\section{Introduction}
\label{sec:intro}
Blockchain is a disruptive technology for provenance and traceability in supply chains\citep{IBM_consumers,Del_report,IBM_SC,prov}. Provenance refers to the data related to the origin of products and the primary producers, whereas, traceability refers to the trail of ownership and state of the products as they make their way through the supply chain life cycle, i.e., from primary producers to the end-users. However, providing traceability and protecting privacy are two conflicting goals in blockchain enabled supply chains\citep{uesugi2020short,mitani2020traceability}. Logging traceability and provenance data on blockchain may pose risks, such as exposure of trade secrets for many supply chain participants. The specific information that constitutes trade secrets may vary across supply chain domains, but generally includes any piece of information that could offer competitive advantage such as product formulas, the location of origin, list of suppliers and payment contracts. Nevertheless, the dilemma between privacy and traceability is similar in other blockchain based decentralised applications.\par

Most blockchain based traceability frameworks proposed in the literature \citep{malik2018productchain, tsang, wu8846964} and industry solutions \citep{IBM_SC,prov} rely on permissioned blockchains, wherein information is only revealed to authorised participants or validating entities. For instance, in consortium blockchains, where peer nodes represent several organizations, a common ledger is shared among the peers of the participating organizations. The ledger provides an audit trail of supply chain events, thus affording provenance and traceability of products. 
Since the events are associated with the digital identifier of supply chain participants, any peer accessing the ledger has full information about the product, its trade flows, or the participants related to the supply chain activities. \par

To address this data privacy problem, data can be encrypted before sharing it on the ledger~\citep{manzoor2020proxy, guo2020data}. However, encrypted data cannot be used in systems which use blockchain technology for instant verification of provenance. Some solutions\citep{li2018fppb, gabay2020privacy} propose to record supply chain event data anonymously to protect trade flows. However, reliance on known identifiers is necessary to allow authorised participation and to have accountability and trust mechanisms integrated with blockchain \citep{malik2019trustchain, volkanTrust2019}. Another solution proposed by Hyperledger Fabric\citep{hyperledger_private}, a permissioned blockchain platform, is the use of private channels for privacy preservation. Although, this limits the visibility of supply chain events within a channel, it is inflexible to support selective data protection by a peer and involves additional overheads \citep{hyperledger_private}. 

A reliable solution would be not to share the private data itself but to generate and share computations\citep{pennekamp2020private} and proofs on the data\citep{jiang2019privacy} in line with the system's traceability and provenance criteria.
However, a blockchain based privacy preserving mechanism employing data computations and proofs, would require additional resource consumption from supply chain participants. Unless there are monetary incentives to compute and share privacy-preserved provenance and traceability information, adoption of blockchain technology in supply chain will remain undervalued. Although  privacy preservation coupled with monetary incentives can encourage more participation from supply chain entities, devising such a mechanism is challenging. The reason is that the blockchains that underpin cryptocurrency cannot directly be used for physical asset transfers such as in supply chains, and in absence of cryptocurrency it is hard to ensure incentive amounts are both kept private and paid as agreed.

\begin{figure}[t]
  \centering
  \includegraphics[width=\linewidth]{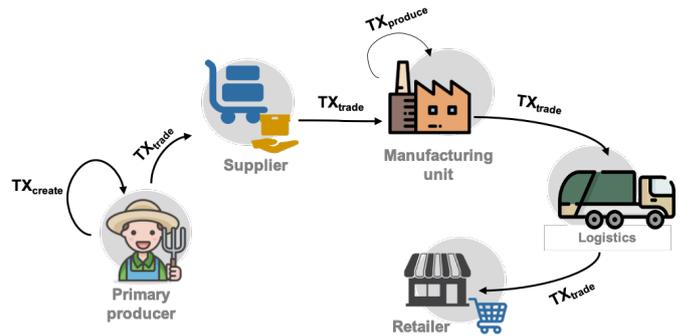}
  \caption{Transaction vocabulary with supply chain process}
  \label{fig:sc}
\end{figure}

In this work we propose privacy preservation and incentive enforcement mechanisms on permissioned blockchain based on Zero Knowledge Proofs (ZKPs) (see Section \ref{sec:premzkp}) and commitment schemes (see Section \ref{sec:prempcs}). ZKPs help to prove possession of a secret without revealing the actual information whereas commitment schemes can be used to verify a committed secret later in time. Supply chain participants can provide ZKP proofs and get reciprocated by the committed incentive amounts for utilizing their resources. The blockchain can verify these proofs, initiate an off-chain payment mechanism and log the results in an immutable way. 
PrivChain leverages the blockchain based supply chain model proposed in \citep{malik2018productchain} and described in Figure~\ref{fig:sc}. The supply chain participants register products on the ledger with product specific data and events. IoT sensor data, such as location, quality, and temperature are hashed and linked to the transactions that log product's events. Availability of this data contributes to the provenance and quality of the product.  
We propose that instead of sharing this data, participants share proofs of their valid data pertaining to products.
 The verification of such proofs is then automated by a blockchain smart contract, which is a self executing software program invoked when some pre-defined conditions are met. The successful verification attests to the claim made regarding quality or provenance of a product. The verification results are then made available with the final product. Sharing proofs rather than the supply chain data allows supply chain entities to keep product and trade specific information private. Since the proof generation incurs additional effort for the supply chain participants, a monetary incentive mechanism is proposed to reward participants for generating successfully verified proofs. PrivChain framework is applicable to other use-cases, such as energy trading and in healthcare where sharing data on blockchain compromises privacy yet it is essential for accountability.

\smallskip

The main contributions of our work are:
\begin{itemize}
\item We devise a ZKP based privacy preservation solution for blockchain enabled supply chains, where the participants provide proof of provenance claims without disclosing privacy-sensitive data such as location and trade flows.
\item We design a smart contract to automate the verification of the provenance proofs and the integration of the incentive mechanism that enforces instant payment of rewards from buyers to sellers for providing verified proofs. 
\item We implement the proposed framework on Hyperledger Fabric and demonstrate that the overheads for proof verification are minimal.
\end{itemize}
The rest of the paper is organized as follows. Section 2 discusses relevant background knowledge. Section 3 details PrivChain architecture and the related components. The performance evaluation and security analysis is outlined in Section 4. Lastly, Section 5 concludes the paper.

\section{Preliminaries}
\label{sec:prem}
In this section, we present a brief description of the supply chain use case and the transaction vocabulary~\citep{malik2018productchain} which will be used as a baseline for our proposed work. Next, a background of ZKP schemes is given with a brief discussion on our choice of ZKP variant for masking provenance data. In addition, we present a brief description of the Pedersen commitment scheme, which provides the basis for our incentive mechanism for providing proof of provenance. 
\subsection{Supply Chain Use Case and Transaction Vocabulary} \label{sec:premsc}
A blockchain-enabled supply chain model, as shown in Figure~\ref{fig:sc}, consists of producers, suppliers, manufacturers, logistics, retailers and consumers who store the physical supply chain events such as production and trade on the blockchain. To demonstrate our privacy preservation mechanism, we consider a wine supply chain, where the wine producer provides a proof for the grape production region but does not want to reveal the exact location of production (to keep his suppliers in that region as a trade secret). Thus, we provide location proofs rather than sharing exact locations for provenance verification.
The proposed underlying network is built on multiple regional side chains \citep{malik2018productchain} that operate in parallel to increase scalability. Each of these side chains is a permissioned blockchain network with a dedicated administrator and a certification authority for registering participants.
To participate in the network, a supply chain entity first registers with the certification authority of his region and obtains a blockchain identity which certifies his digital profile on the ledger. The supply chain participants log supply chain events on the blockchain using the transaction vocabulary proposed in \citep{malik2018productchain}.
For the wine supply chain use case, a grape producer registers a new batch of commodity on the ledger using a create transaction, $TX_{create}$. When the grapes are sold to an intermediate supplier or a wine producer, the trade is recorded using a trade transaction, $TX_{trade}$. A wine producer processes the grapes sourced from multiple suppliers and registers the final product (e.g., a bottle of wine) on the blockchain using a produce transaction, $TX_{produce}$, which also contains the identifiers of constituting batches of grapes using $TX_{create}$. The final product, a wine bottle, is then traded through logistics and retailers to the end consumers.
The wine producers prove origin information through blockchain, without revealing the exact location and identities of grape producers. 
The grape producers provide location proofs using ZKPs. The wine producers on the other hand commit to incentive payments to grape producers and keep the sellers' information private in $TX_{produce}$. Without loss of generality, we limit the seller and buyer model to grape and wine producers only and eliminate any intermediate suppliers.

\subsection{Zero Knowledge Proofs}\label{sec:premzkp}
ZKPs are cryptographic methods in which a prover can attest to a verifier that some secret information is true without revealing any details about the secret. 

Depending on the requirement of interactions between the prover and the verifier, ZKPs are interactive or non-interactive. For a distributed network such as blockchain, interactive ZKPs are impractical due to the communication overhead for verification. Zero-Knowledge Succinct Non-Interactive Argument of Knowledge (ZK-SNARKs)~\citep{snark_gentry} are an efficient non-interactive variant of ZKPs. In addition, the proofs generated are of constant size despite the complexity of proof generation. One practical example of the use of ZK-SNARKs is Zcash \citep{hopwood2016zcash}. The steps involved in the process of non-interactive ZKPs are defined below:

\begin{itemize}
\item \textbf{Setup:} This step is performed by a trusted third party and entails key generation. The key generator \textit{G}, takes a secret parameter $\lambda $ to generate two publicly available keys: proving and verification keys. 
\item \textbf{Proof Generation:} A prover uses the proving key, a common input and a secret, $\delta$ to generate a proof. 
\item \textbf{Verification:} The verifier runs a function that takes proof, common input and the verification key as inputs and returns true if the proof is correct and false otherwise.
\end{itemize}

Generic ZKPs, e.g., ZK-SNARKs, are less efficient than specific ZKPs such as ZK Set Membership (ZKSM) or ZK-Range Proofs (ZKRPs) \citep{koens2018efficient}. The ZKSM scheme allows anyone to prove that a secret $\delta$ lies within a given set $[u,v]$. ZKP-Range Proofs (ZKRPs) are a special instance of ZKSM \citep{koens2018efficient} , which prove that a certain secret number lies within a specific numeric range.  To create ZKRP, the secret $\delta$ is decomposed into base $u$ using

\begin{equation}
  \delta = \sum_{j=0}^{l} \delta_j u^j
\end{equation}

if each $\delta_j$ belongs to the interval 
$[0, u)$, then $\delta \in [0,u^l]$. The ZKSM algorithms\citep{morais2019survey} can be easily
adapted to carry out ZKRP computation as it is specific to numeric intervals as compared to ZKSM. When compared to a generic ZKP, ZKRP is proven to reduce the computation overhead of proof verification by an order of 10 \citep{koens2018efficient}. Hence, ZKRP can be applied to  a range of decentralised applications where numerical data needs to be proven without actually revealing it such as account balances (e-finance), sensor readings (IoT), electronic auctions (e-commerce), rating/trust scores (supply chains), and distribution of sales (energy trading).

Bulletproofs\citep{bunz2018bulletproofs} are short and efficient ZKRPs designed specifically for blockchain. They have a performance advantage as the size of proofs are logarithmic to the input size and they do not require a trusted setup. Note that, the choice of the range proof methods differs based on the computation power of the prover and the verifier. 
Bulletproofs with a smaller proof size reduce the transaction size when compared to other variants of ZKRP such as signature based methods\citep{camenisch2008efficient}. However, signature based methods incur lower verification complexity only when the range size is less than $10^2$ bits\citep{canard2013new}. For the proposed privacy preservation scheme, we choose non-interactive signature based range proofs \citep{camenisch2008efficient} based on 1) our reliance on permissioned blockchains for trusted setup and verification whereby the computation overhead for setup and verification is delegated to the blockchain provider; 2) the lower proof verification complexity of non-interactive signature based range proofs given the high transaction send rate in typical supply chain use cases.
 \begin{figure*}[h]
  \centering
  \includegraphics[width=11cm]{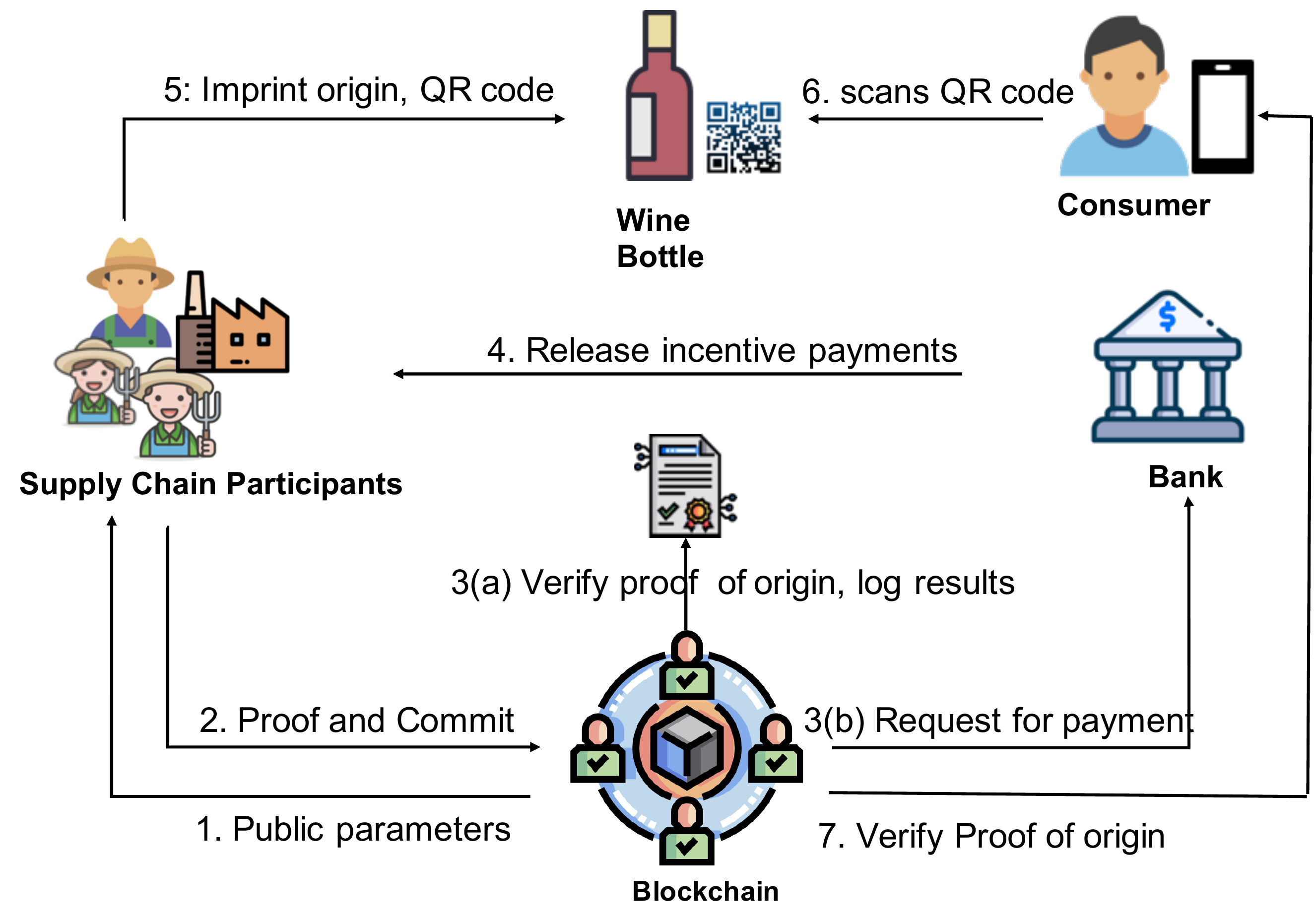}
  \caption{Overview of Workflow}
  \center \label{fig:framework}
\end{figure*}
\subsection{Pedersen Commitment Scheme}\label{sec:prempcs}
Pedersen commitment \citep{pedersen1991non} is a non-interactive secret sharing scheme. It allows a user to commit to a secret, and send a verifier a commitment to the secret without disclosing the secret itself. The secret is shared with the verifier at some later point in time. The verifier computes a new commitment from the secret and verifies it by comparing it to the previous commitment shared by the committer. This scheme is based on the following three steps:

\begin{itemize}
\item  \textbf{Generator Setup:} Let $p$ and $q$ denote two large prime numbers such that $q$ divides $p - 1$. $G_q$, a unique subgroup of $\mathbb{Z}_p$ of order $q$, is chosen by a trusted third party where $g$ is the generator of $G_q$. Next, $h \in G_q$ is chosen such that $log_g h$ is a secret. The parameters $p$ and $q$ are then made public by the trusted third party. 
\item \textbf{Commit:} To commit to a secret message $m$, first a random number $r \in \mathbb{Z}_p$ is chosen. Then, a commitment at the committer's end is computed as $com_c =g^m h^r$ mod  $q$ and sent to the verifier at time $t_m$.
\item \textbf{Commit Open:} This step involves the commitment verification at time $t_n$ where $n > m$. The committer reveals $m$, and $r$ to the verifier. The verifier computes another commitment $com_v$ and compares it with the previous commitment to validate if and only if $com_c \stackrel{?}{=} com_v$.
\end{itemize}

We choose this scheme based on its strong properties, namely: 1) it does not allow a committing party to change the $m$ once committed and 2) it is hard for an unbounded adversary (in terms of time and resources) to find $m$ \citep{pedersen1991non}.

\section{Privacy Preservation in Blockchain-Enabled\\ Supply Chains}
\label{sec:arch}
Recall from  Section~\ref{sec:premsc}, that our aim is to preserve the privacy of the provenance data and trade flows associated with a final product, while ensuring that the primary producers receive the pre-agreed incentives for providing provenance proofs. In this section, we present an overview of PrivChain and provide details of the components involved in recording provenance of the grapes on the blockchain, protecting trade flows of wine producers, and actioning the incentive payment mechanism. Finally, we outline how end-consumers can query the blockchain to verify the origin of grapes used in production of a wine bottle.

\begin{figure*}[h]
  \centering
  \includegraphics[width=14cm]{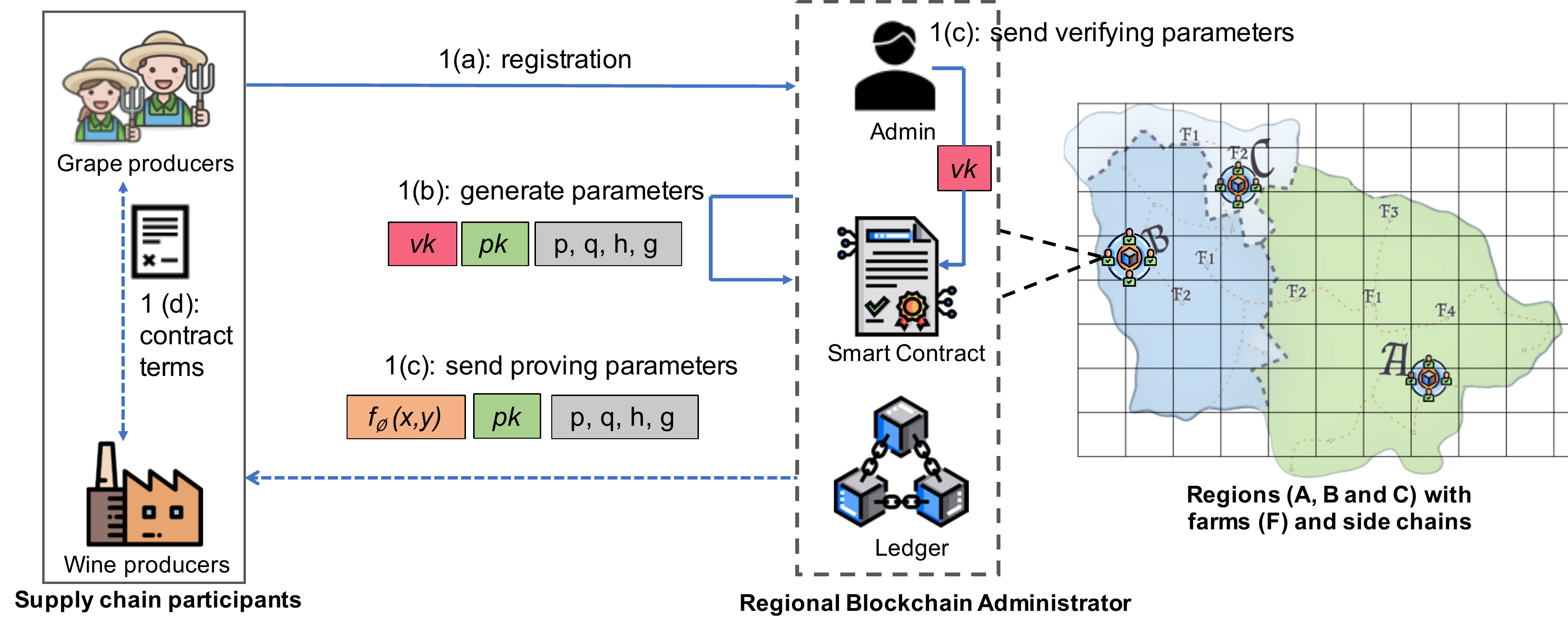}
  \caption{Setup Phase}
  \label{fig:setup}
\end{figure*}

\subsection*{Overview}

The high-level workflow for PrivChain is shown in Figure~\ref{fig:framework}. The blockchain service provider sends the public setup parameters to the registered supply chain participants (Step 1). Using these parameters, the grape producers generate ZKRP location proofs, $\phi_{loc}$ (see Section~\ref{sec:premzkp}), to register a new batch of grape commodity using $TX_{create}$. The grape producer also generates a commitment $com_{inc}$ to the pre-agreed incentive amount, $\$_{inc}$ (see Section~\ref{sec:prempcs}), using $TX_{trade}$. Both the $\phi_{loc}$ and $com_{inc}$ are recorded on the blockchain through the initiation of a $TX_{trade}$ by the grape producer (Step 2). Next, a verification and incentive smart contract, $VISC$, verifies $\phi_{loc}$ and registers the grape commodity along with the verified region (Step 3(a)). It also generates a request for incentive payment to a financial institution, such as a bank using $Req_{pay}$ (Step 3(b)). The bank verifies $com_{inc}$ and releases the incentive amount to the grape producer (Step 4). In the final stage of manufacturing, the wine producer encrypts the grape producers' and commodity's details and stores the identifier of a final product on the ledger using $TX_{produce}$. The wine producers then imprint the origin information on the wine bottle along with a QR code which can be used to query the blockchain (Step 5). The end-consumers can scan the QR code (Step 6) to retrieve the origin information from the blockchain which can be used to verify the printed origin on the wine bottle (Step 7). Having covered the high level workflow, we discuss each of the steps in more detail next.

\subsection*{Step 1: Setup Phase}
The setup phase is illustrated in Figure~\ref{fig:setup}. Recall from Section~\ref{sec:premsc}, we rely on a scalable network model where multiple regional side chains are operating in parallel. Figure~\ref{fig:setup}, for example, shows a map with different regions, each depicting a side chain. We describe the specifics within one side chain, which is replicated across all side chains. In order for a  grape producer to generate $\phi_{loc}$ and $com_{inc}$, public parameters for setup and commitment generation are required (as outlined in Section~\ref{sec:premzkp} and Section~\ref{sec:prempcs}).
For a particular side chain, a blockchain administrator generates the following standard ZKRP setup parameters \citep{morais2019survey}: proving key $pk$, verification key $vk$, and $f_{\phi}(x,y)$ for generating $\phi_{loc}$ and a proof verification function. The grape producer takes location input from a sensor device in form of a signed commitment to GPS coordinates $(x,y)$. Specific to our use-case, we assume that the GPS device is trusted \citep{ranganathan2016spree} and cannot be tampered with. In case the devices are not guaranteed to be trusted, we can use existing solutions in literature such as signal strength or  frequency analysis to identify spoofed signals, etc \citep{leibner2019gps}. The grape producer uses the signed commitment to $(x,y)$ as an input to $f_{\phi}(x,y)$, which is used to compare the input to authorised locations in a particular region. 

However, computing signatures for every possible GPS coordinate in an authorised region is impractical for a business administrator. Thus, we propose to map the GPS coordinates of a particular region to a Military Grid Reference System (MGRS)\citep{jensen2014national} grid index. Using MGRS, Earth's surface can be divided into squares of 1000m × 1000m, 100m × 100m, 10m × 10m or 1m × 1m. Any specific point on Earth can then be represented as an area specified by grid indices. For our use-case, a region can be represented by a grid where each square is of size 10m × 10m (across all the side chains).  The blockchain administrator can pre-compute the signatures on MGRS grid indices of known farms within each square in a grid. However, during the process of generating $\phi_{loc}$, conversion of farm coordinates into grid indices will also be required which can be later mapped to the authorised region specific to each side chain. The region boundaries and names can be agreed upon by the side chains' administrators, such that each $\phi_{loc}$ can be mapped to a specific region name.\par

\begin{algorithm}[!b]
\label{algo:zkrp}
\caption{Non-Interactive ZKRP Generation for Location Proof}
\begin{algorithmic} [1]
\REQUIRE Request to log $TX_{cr}$ from the seller, location $\delta = (x,y)$ signed by the GPS device at the farm
\STATE The seller wants to keep $\delta$ secret and prove that it lies within MGRS indices of registered farms in a particular region.
\FOR{\textbf{each} $a \in \delta$} 
\STATE the seller generates a commitment (as described in Eq. 1), based $y=g^a$ to the base $g$ where $a \in [0,u^l]$
\STATE the seller chooses a random $v_j \in \mathbb{Z}_p$ for every $j$ in Eq. 1 and computes $t_j = g^{v_j}$.
\STATE The seller computes a commitment (see Algorithm 15 \citep{morais2019survey}) using a cryptographic hash function, $com_a = HASH(g,y,t)$.
\STATE The $com_a$ and range for authorized region $[0, u^l]$ is given as an input to a proof generating function, $f(a)$. 
\STATE For each $j \in \mathbb{Z}_l$,  $f(a)$ computes $r_j = v_j - ca$ and checks if $g^{r_j}y^c \equiv t_j $, it generates a witness, $w$. 
\STATE The witness, $w$ is then signed by the seller's private key, $pk$ to generate $\phi_a$.

\ENDFOR
\end{algorithmic}
\end{algorithm}

\begin{figure*}[t]
  \centering
  \includegraphics[width= 11.5cm]{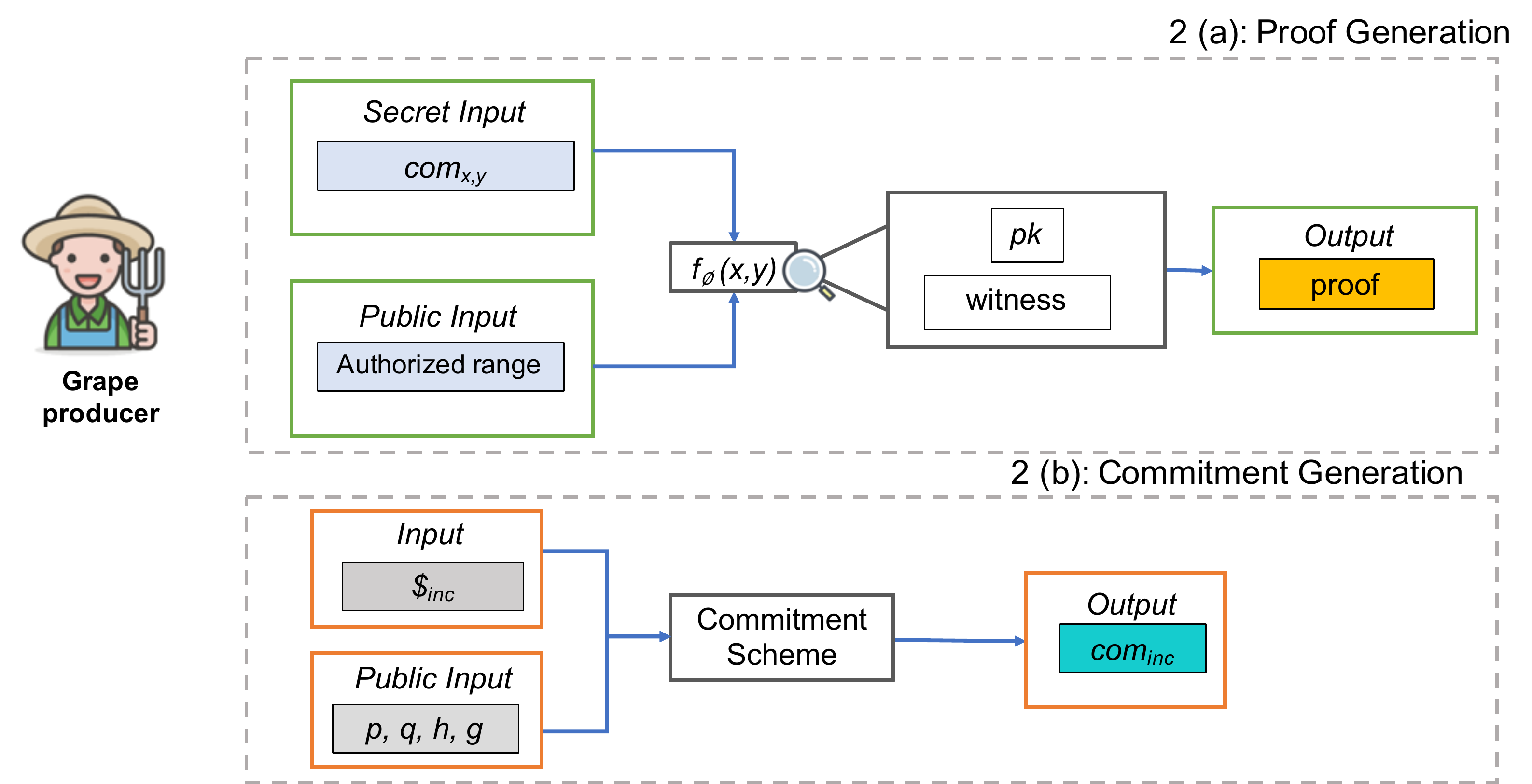}
  \caption{Proof and Commitment Generation}
  \label{fig:proofGen}
\end{figure*}
During the setup phase, the blockchain administrator also provides the public parameters $p$, $q$, $g$ and $h$ for generating commitments such as $com_{inc}$. The commitment and proof generating parameters are then distributed to the registered supply chain participants to generate $\phi_{loc}$ for their commodity and $com_{inc}$ for protecting incentive amounts.

\subsection*{Step 2: Proof and Commitment Generation for ZKRP}
The proof generation is executed off-chain at the grape producer's end and illustrated in Figure~\ref{fig:proofGen}. To create $\phi_{loc}$, a grape producer inputs the secret location coordinates $(x,y)$ to a proof generating function, $f_{\phi}(x,y)$. For our use-case, the location data $(x,y)$ is provided by a GPS sensor registered at the grape producer's farm. The sensor device signs the location data using its private key. We assume that the public keys of the GPS devices are known to the blockchain administrator. The proof generation steps can be summarized as (for details of the signature-based range proof construction, see \citep{morais2019survey}):
\begin{itemize}[leftmargin=*]
    \item Generation of the digital signatures for the authorised region: To construct a proof, a grape producer uses $f_{\phi}(x,y)$ to map $(x,y)$ to an authorised region. The signed commitments for all $(x,y)$ in the authorised region are provided by the blockchain administrator which are used as the public input for $f_{\phi}(x,y)$. In the case of MGRS mapping (see step 1: setup phase), the public input will comprise of commitments to MGRS indices of registered farms only.  
    \item Generation of the commitment for the secret $(x,y)$: To generate a commitment $com_{x,y}$ for $(x,y)$, the grape producer first decomposes the secret $(x,y)$ into \textit{base-u}, such that each element belongs to the interval $[0,u)$. The producer then computes digital signatures on each element and blinds the signatures by raising them to a randomly chosen exponent $v \in \mathbb{Z}_p$, such that it becomes computationally infeasible to determine the signed elements.
    \item Generation of the proof:
    When a producer has a valid $com_{x,y}$ with respect to the authorised region, a \textit{witness} parameter is generated. In the final step, the grape producer signs the \textit{witness} using $pk$ to generate $\phi_{loc}$, which is the final output of $f_{\phi}(x,y)$.

\end{itemize}

\begin{algorithm}[b]
\caption{Commitment Generation for the Incentive, $\$_{inc}$}
\begin{algorithmic} [1]
\REQUIRE valid $TX_{trade}$ parameters, negotiated $\$_{inc}$ between a buyer and a seller, $p$, $q$, g (generator of $G_q$)
\STATE The seller chooses a random $r \in \mathbb{Z}_p$ and $h \in G_q$
\STATE The seller then computes the commitment to incentive amount, $\$_{inc}$, using $com_{inc} =g^{\$_{inc}} h^r$ 
\STATE The $com_{inc}$ is then included in $TX_{trade}$ by the seller before it is logged on the ledger. 
\end{algorithmic}
\end{algorithm}

The non-interactive ZKRP used for generating $f_{\phi}(x,y)$ is outlined in Algorithm 1.  It is important to note that $f_{\phi}(x,y)$ cannot be altered by the grape producer or any other unauthorised participants.The grape producer registers a new batch of grapes using $TX_{cr}$. The structure of $TX_{cr}$ is given below:

\begin{equation}
    TX_{cr}= \Big[\;ID_g\;|\;H_{data}\; |\;L_{\phi_{loc}}\;|\;Sig_{s}\;\Big]  
\end{equation}
where $ID_g$ is the identifier of the grape commodity, $H_{data}$ is the hash of the commodity data (e.g. commodity type, quantity, unit price, etc.), $L_{\phi_{loc}}$  is the link to $\phi_{loc}$ which has been generated using $f_{\phi}(x,y)$. $PU_s$ is the public key of the seller and $Sig_{s}$.\par

Next, the grape producer generates a commitment, $com_{inc}$. Prior to a supply chain trade event, there is a negotiation between the seller (grape producer) and the buyer (wine producer) to set the incentive amount $\$_{inc}$, which could be included in a digital contract related to the commodity trade. However, this exposes $\$_{inc}$ to the blockchain network. On the other hand, if $\$_{inc}$ is encrypted, blockchain cannot verify $\$_{inc}$. To keep $\$_{inc}$ confidential and to have the wine producer adhere to $\$_{inc}$, a commitment $com_{inc}$ is generated by the grape producer using the Algorithm 2. This commitment $com_{inc}$ attests to the incentive amount $\$_{inc}$ which the producer is expecting in return for providing a valid location proof $\phi_{loc}$.
$\phi_{loc}$ and $com_{inc}$ are then sent to the blockchain during the trade of commodity using $TX_{trade}$. The structure of $TX_{trade}$ is given below:
\begin{multline}
      TX_{trade}= 
      \Big[\:ID_g\:|\:H_{data}\: |\:L_{\phi_{loc}}\:|\:com_{inc}\: \\ |\:region\:|\:Sig_{s}\:|\:PU_s\:|\:Sig_{b}\:|\:PU_b\:\Big]
\end{multline}
where $ID_g$ is the identifier of the grape commodity, $H_{data}$ is the hash of the commodity data, $L_{\phi_{loc}}$ is the link to $\phi_{loc}$, $com_{x,y}$ is the commitment to $\$_{inc}$, $region$ attribute is populated after $\phi_{loc}$ has been verified, $Sig_{s}$ and $PU_s$ are the signatures and public key of the seller, and  $Sig_{b}$ and $PU_b$ are the signatures and public key of the buyer.\par

The proof verification process and incentive request mechanism is described in Step 3(a) and 3(b) respectively.

\subsection*{Step 3(a): Proof Verification}
$TX_{trade}$ invokes a verification and incentive smart contract, VISC which initiates an on-chain proof verification process (see Section~\ref{sec:premzkp}:verification). The verification is done only at the time $TX_{trade}$ is initiated. The contract uses the verification key $vk$ enlisted in its properties against the proving key $pk$ distributed to the registered participants. The verification function returns true if $\phi_{loc}$ is successfully verified, otherwise it returns false. $TX_{trade}$ contains a \textit{region} attribute where the proof verification result is specified. When a valid $\phi_{loc}$ is provided, the name of the region is stored in the \textit{region} attribute. In case $\phi_{loc}$ is invalid, the \textit{region} attribute is populated as ``not verified". 
  Following the verification of $\phi_{loc}$, VISC transfers the ownership of grape commodity to the wine producer and generates a request to a bank for incentive payment, $Req_{pay}$ (discussed in Step 3(b)).\par
   To keep blockchain usage open for all registered participants, provision of proofs is not mandatory for the trade of commodity. If no $\phi_{loc}$ is provided during the trade, ``proof not provided" is embedded in \textit{region} attribute of $TX_{trade}$ and $Req_{pay}$ is not issued. However, based on the supply chain use-case, the provision of $\phi_{loc}$ can be made mandatory.

\subsection*{Step 3(b) and Step 4: Incentive Payments}
\begin{figure}[t]
  \centering
  \includegraphics[width= \linewidth]{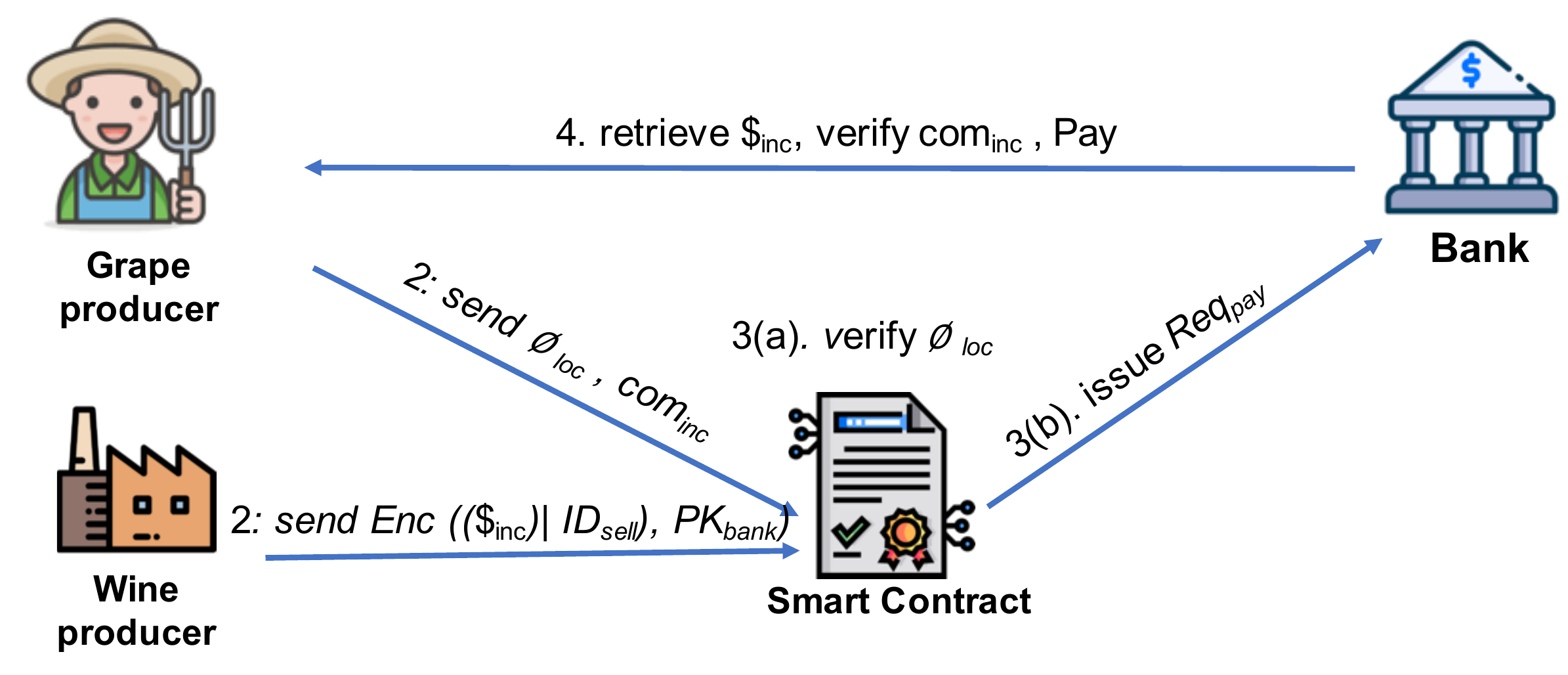}
  \caption{Incentive payment: request generation and verification}
  \label{fig:ver}
\end{figure}

The incentive mechanism is automated through smart contracts to ensure participants contributing valid data, i.e., verified proofs, get instant returns from the trading parties. Furthermore, the incentive amounts are kept private from the blockchain network.
As shown in Figure~\ref{fig:ver} and Eq. 3, during the $TX_{trade}$, the seller also sends $com_{inc}$ (generated in Step 2) to VISC. The buyer on the other hand encrypts the incentive amount $\$_{inc}$, and seller's ID, $ID_{sell}$ using the bank's public key $PK_{bank}$, signs it using $Sig_{buy}$ and sends it to VISC. Then, the VISC generates $Req_{pay}$ for the bank notifying a pending incentive payment as:
\begin{equation}
    Req_{pay} = \Big[ \: com_{inc}\: |\:
    Enc\:( (\$_{inc}\:|\:r\:|\: ID_{sell}\:), PK_{bank})\: |\:Sig_{buy}\: \Big]
\end{equation}

 After receiving $Req_{pay}$, the bank decrypts $\$_{inc}$ and $ID_{sell}$ using its private key. Next, the bank computes a new commitment $com_{inc}'$  using the Algorithm 3. The bank then compares it with $com_{inc}$ which was issued by the grape producer. If both commitments match, it confirms that $\$_{inc}$ is the agreed incentive amount, and must be paid to the seller for providing a valid $\phi_{loc}$. The bank transfers $\$_{inc}$ to the seller's account. The generation of ${com'_{inc}}$ and comparison with $com_{inc}$ is carried off-chain to reduce any on-chain computation overheads. Once the payment request is processed, a transaction is logged on the ledger confirming the payment. To preserve the privacy of the seller and the buyer, the transaction only includes an identifier for $Req_{pay}$ and the status of payment.
\subsection*{Step 5: Protection of Trade Flows}
\vspace{0.3mm}Recall from Section~\ref{sec:premsc}, the provenance of a wine bottle on blockchain is computed by querying the transactions chained to the grape identifiers involved in $TX_{produce}$. The inclusion of grape identifiers help in locating $TX_{create}$ of each commodity listed in $TX_{produce}$ (see Section ~\ref{sec:premsc}). Using $TX_{create}$, all the relevant details regarding origin, primary producer, etc. can be derived. The grape identifier can also be used to trace all the trade transactions $TX_{trade}$ involved. Hence, knowledge of grape identifiers may lead to the exposure of trade flows involved in the life-cycle of a commodity. Even if the trade flows are hidden from end-consumers, they are accessible to the peers of the blockchain network. To preserve the privacy of these trade flows within the blockchain network, we hide the identifiers of grapes commodities that form the raw material for the wine in a bottle.    

\begin{algorithm}[t]
\caption{Commitment Verification}
\begin{algorithmic} [1]

\REQUIRE valid $TX_{trade}$, valid $\phi_{loc}$, $g$ (generator of $G_q$)
\STATE The  verification and incentive smart contract, VISC generates a $Req_{pay}$ according to Eq. 2. 
\STATE the bank decrypts $\$_{inc}$ and $r$ from  $Req_{pay}$
\STATE The bank then computes another commitment ${com'_{inc}}$ based on the decrypted values using ${com'_{inc}} = g^{\$_{inc}} h^r$.The bank then checks: 
\IF {$com'_{inc}$= =$com_{inc}$} 
\STATE The incentive amount, $\$_{inc}$ is paid to the seller's account which is identified using $ID_{sell}$ in $Req_{pay}$
\STATE The bank logs a payment transaction on the ledger
\ELSE 
\STATE the bank logs a dispute against the commodity identifier, $CID$.
\ENDIF \\
\end{algorithmic}
\end{algorithm}

 As described in Step 3(a), once $\phi_{loc}$ is verified, the verified region name is stored in $TX_{trade}$. When $TX_{produce}$ logs a wine bottle on the ledger, all the pre-verified regions are included in \textit {regions} attribute of $TX_{produce}$ in a readable form, while the identifiers of grape commodities are stored in encrypted form to preserve the privacy of the producers and the suppliers involved in the production of the wine bottle. Upon provenance query from an end-consumer, only the names of the pre-verified regions from $TX_{produce}$ are returned. The modified $TX_{produce}$ is defined as:
\begin{multline}\label{eq:produce}
    TX_{produce}  =  \\  \Big[ ID_{FP} \: |  \: Enc(\:(ID_{g1},\hdots, ID_{gn}), Key) \: | \: regions\:|\: Sig_{buy}\Big]
\end{multline}

where $ID_{FP}$ is the final product identifier for the wine bottle, $ID_{gi}$ is the identifier for grape commodity $i$ encrypted using a symmetric encryption key, $Key$, and $Sig_{buy}$ is the signature of the buyer (wine producer). Note that $Key$ is only distributed to the authorised participants who need to verify the constituting identifiers in a wine bottle, for example, a food regulatory authority which may use blockchain transactions for verification of provenance claims. The key distribution problem is out of the scope of this work. However, some external nodes can be appointed for managing keys such as ``access guards" proposed in \citep{pennekamp2020private}.
\subsection*{Step 6 and 7: End-Consumer Query}\label{sec:query}
Once the wine bottle is produced, it is then traded to a retailer for selling. The wine bottle has printed provenance information and a QR code. The QR code enables an end-user to generate a query transaction using a blockchain consumer application. The query transaction returns the \textit {regions} attribute from $TX_{produce}$ for the  $ID_{FP}$ of the wine bottle specified in QR code. End-consumers can compare the results returned through the blockchain application with those printed on the bottle. 

\section{Evaluation and Results}
In this section, we first present the business model formulation in the context of Hyperledger Fabric followed by the experimental setup. Next, we present results quantifying the performance of our system for relevant benchmarks. Finally, a qualitative security and privacy analysis of PrivChain is presented.

\subsection{Business Model}
The proposed system is implemented on Hyperledger Fabric\footnote{https://www.hyperledger.org/use/fabric}, an enterprise grade permissioned blockchain platform created for integrating blockchain with business applications. For evaluation of PrivChain, we devise a commodity trading business network in Hyperledger Fabric comprising of: 
\begin{itemize}
\item Participants: include grape producers, a wine producer, a bank and an end-consumer. 
\item Assets: we define two assets for our business model; commodity (grapes) and a final product (wine bottle). 
\item Chaincode (smart contract): consists of functions which initiate a ledger, query or update it. These chaincode functions include registering a grapes produce ($TX_{create}$), registering a wine bottle ($TX_{produce}$), trading an existing produce or bottle ($TX_{trade}$), and querying a wine bottle ($TX_{query}$) according to the transaction vocabulary described in Section~\ref{sec:premsc}. For our proof of concept, we deploy a single chaincode which also embeds the corresponding functions of verification of proofs and request generation for incentive transfer.
\end{itemize}
\subsection{Experimental Setup}
The deployment of the business network and performance tests are carried out on a Dell Notebook (Intel Core i7, 2.21 GHz, 8 GB memory). A complete business network setup for our proposed system involves three components: Hyperledger Fabric for chaincode deployment, Hyperledger Fabric network model and Hyperledger Fabric Software Development Kit (SDK) for client application interaction with the chaincode.\par 
We build a Fabric network of two organizations, depicting grape producers (sellers) and wine producers (buyers). Each organization consists of two peer nodes, one orderer (using SOLO as the ordering method),  a database (goleveldb) and a certification authority. Before a client application can interact with the Fabric network, it must be authorised by the respective certification authority. Once authorised, the client application can invoke or query from the ledger. \par
In addition, for the implementation of ZKRP \citep{camenisch2008efficient} and Pedersen commitment, we choose the ING-bank repository\footnote{https://github.com/ing-bank/zkrp}, an open-source reusable library for creating and verifying ZKRPs and set membership proofs. The commitments $com_{inc}$ and proofs are generated on the client application. However, verification is done on-chain, using the verification key, $vk$ provided by the business network administrator. The chaincode generates $Req_{pay}$ using Hyperledger chaincode events and sends them to the respective bank's application. Event handling is carried out using Hyperledger Fabric SDK which provides a fabric network package to listen to chaincode events when a specific transaction is executed. For encrypting trade flows as outlined in Step 5, we use npm
Crypto-JS package\footnote{https://www.npmjs.com/package/crypto-js} with AES encryption.
In addition, we used Hyperledger Caliper\footnote{https://www.hyperledger.org/projects/caliper}, a benchmark tool for blockchain performance evaluation. 
The workload of generating transactions registering a wine bottle was distributed across three processes capable of simulating concurrent transactions.

\subsection{Performance Evaluation}
\label{perform}
The proof of concept implementation of our proposed system is evaluated for ZKRP overheads, and registering a wine bottle with encrypted trade flows. ZKRP computations include: setup (by the blockchain administrator), proof generation (by the grape producer), and proof verification (by the chain code). Registering a wine bottle (by the wine producer) involves  a) querying the blockchain for the region information of grape commodities involved, b) encrypting the identifiers of these grape commodities, and c) registering the bottle on ledger.
In the next subsections, we analyse the associated execution times (averaged across 10 runs) in carrying out these computations. The execution times of commitment generation and verification are excluded as these tasks can be accomplished in minimal time.\par

\subsubsection{ZKRP overheads}
\label{zkpoverheads}
\begin{figure}[t]
  \centering
  \includegraphics[width= 8cm]{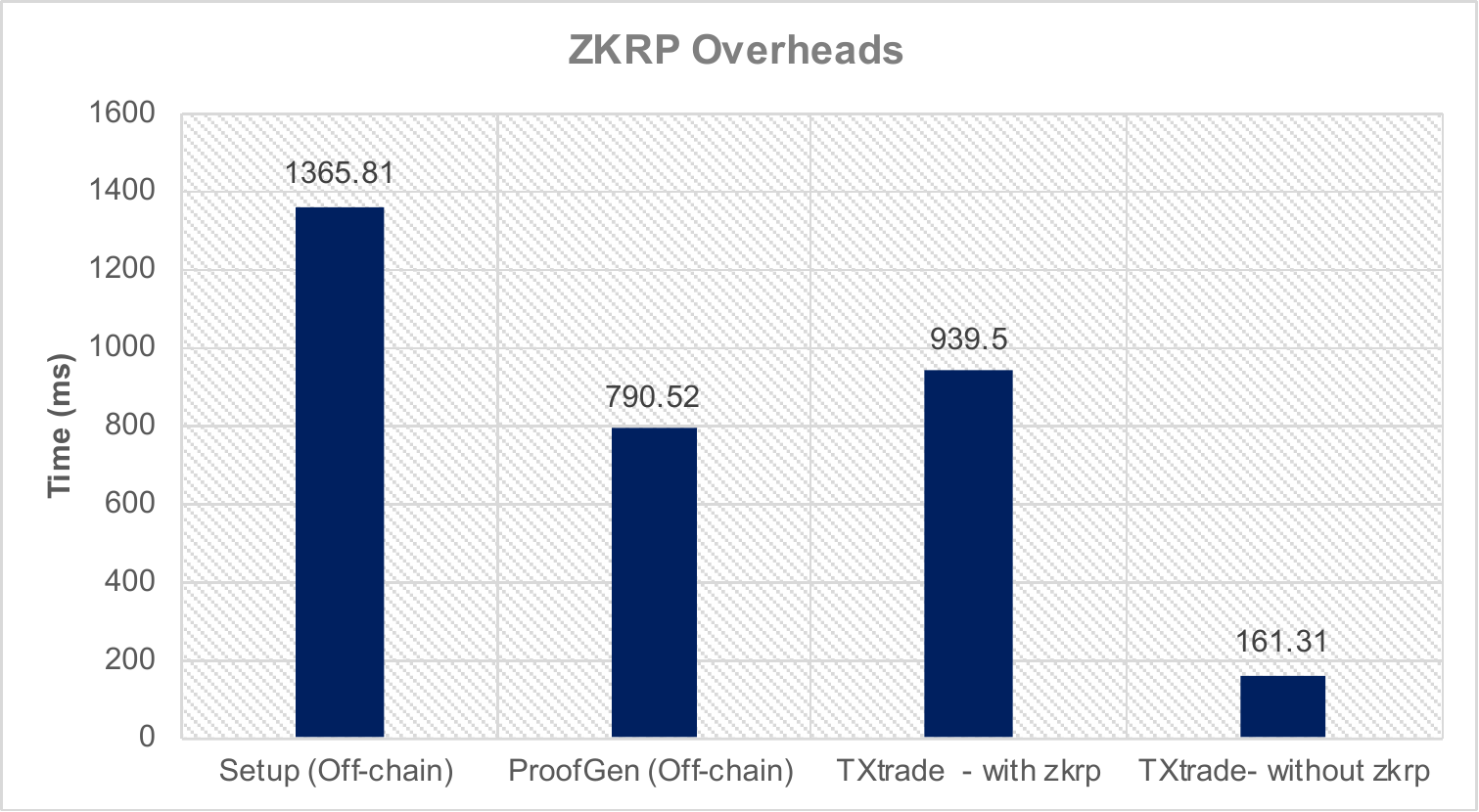}
  \caption{Range proof generation and verification times}
  \label{fig:zkrpOH}
\end{figure}

Figure~\ref{fig:zkrpOH} depicts ZKRP computation times for setup, proof generation and verification (see Section~\ref{sec:arch}). The time required for the setup phase is 1.35 seconds. The setup phase is executed once by a blockchain administrator, thus it is a one-time cost and its effect is minimal. Proof generation takes 0.79s on average which is acceptable given that the proofs will only be generated when a new grape commodity is ready.  The proof generation and storage is carried off-chain. The proofs are not included in blockchain transactions, hence the transaction size is not affected. 
To benchmark the proof verification time, we first define a baseline consisting of a smart contract for trading grape commodities. In the baseline, a smart contract (triggered by $TX_{trade}$) only performs the ownership transfer from a seller to the buyer. In contrast, in PrivChain, VISC not only transfers ownership but also performs proof verification and payment request generation.
As shown in Figure \ref{fig:zkrpOH}, the execution time of $TX_{trade}$ without ZKRP is 161.31 ms whereas for  $TX_{trade}$ with ZKRP it is 939.50 ms. The proof verification overhead compared to the baseline is only 0.77s. 

\begin{figure}[b]
  \centering
  \includegraphics[width=8cm]{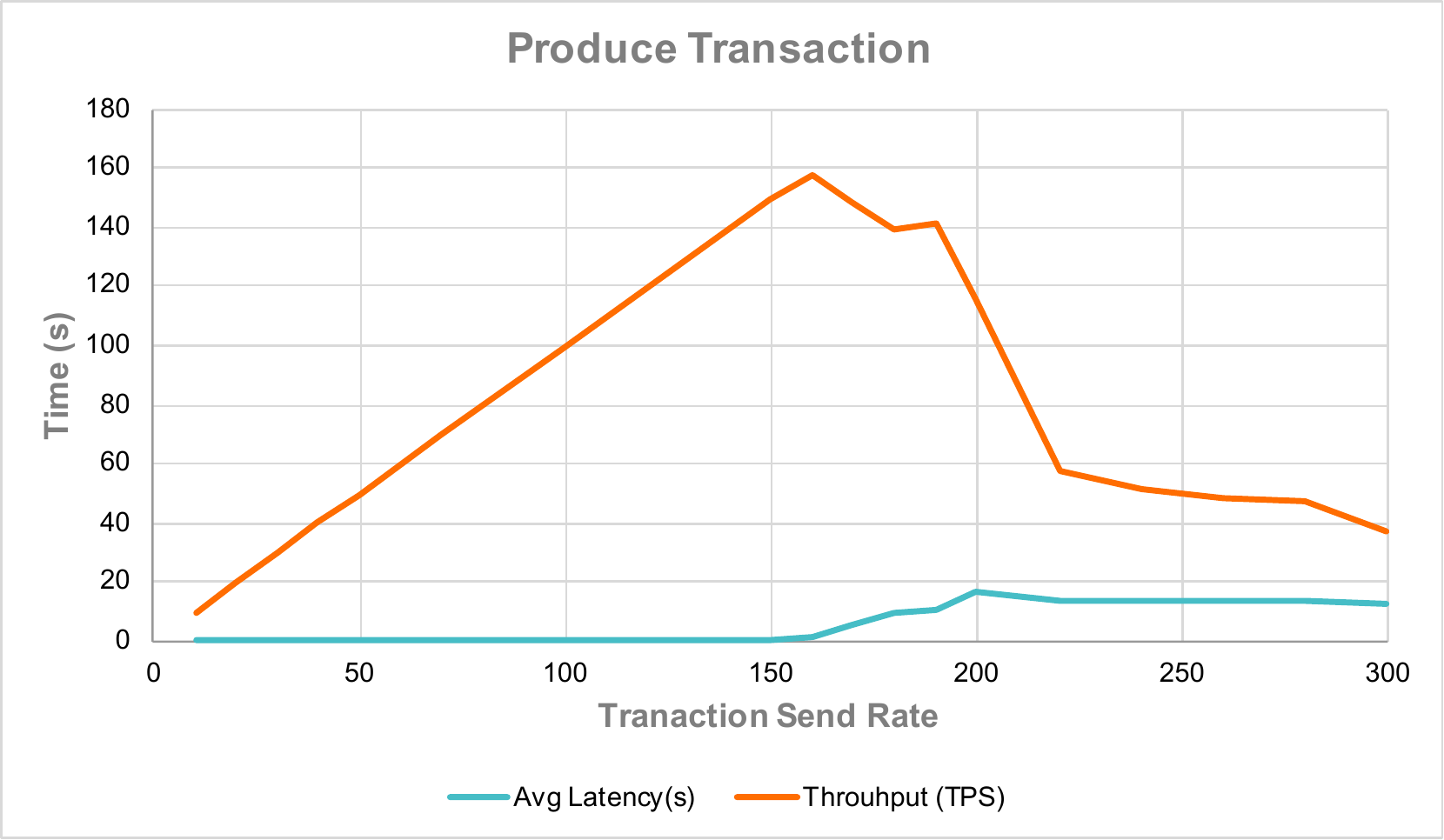}
  \caption{Throughput and Latency of $TX_{produce}$}
  \label{fig:producelat}
\end{figure}

\subsubsection{Registering Products with Encrypted Trade Flows}
\begin{figure}[t]
  \centering
  \includegraphics[width= 8cm]{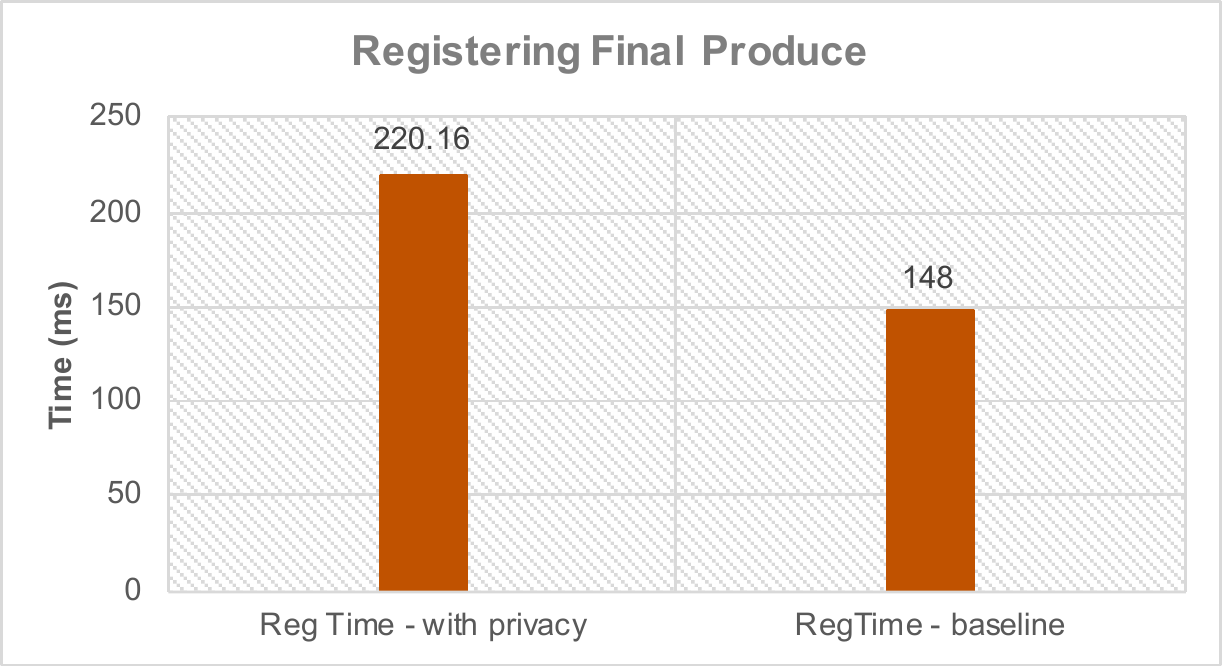}
  \caption{Registering product with encrypted trade flows}
  \label{fig:produceOH}
\end{figure}

 To generate the $TX_{produce}$, the wine producer first queries the blockchain for the \textit{region} attribute of the commodities involved in the production of the wine bottle. The producer then encrypts the identifiers of these commodities, and generates $TX_{produce}$ as defined in Section~\ref{sec:premsc} which registers the wine bottle on the blockchain. Figure~\ref{fig:produceOH} shows that the time required to register a wine bottle (including both the query and encryption steps) is 220.16 ms. The wine bottle registration time for the baseline on the other hand is 148 ms. The results show that querying and encryption of grape identifiers incurs an increase of only 0.07s compared to the baseline. Note that for our proof of concept implementation, we considered three grape commodities involved in the production a wine bottle. If the number of involved commodities is more, the query and encryption time is likely to increase linearly. \par 

Figure~\ref{fig:producelat} presents the throughput and latency of $TX_{produce}$ varying the transaction send rate from 10 tps to 300 tps for the duration of one minute. Throughput is measured as rate at which the transactions are committed to the ledger whereas latency is the time taken from an application sending a transaction to the time it is committed in the ledger. With three concurrent processes, throughput increases linearly reaching a maximum of around 160 tps with the available resources. Beyond this point the network saturates, so the throughput decreases and the latency increases. However, if this trend prevails on a faster computing resource then more validating peers will be required to handle  transaction load higher than 160 tps. 

\subsection{Security and Privacy Analysis} 
Herein, we analyse the major security risks and threats posed by malicious activities with respect to the PrivChain. 

\subsubsection{Linking Attack}
In a permissioned blockchain, participants are registered using permanent identifiers. Recall from Section~\ref{sec:premsc} that a trade transaction includes both the seller and buyer information. 
If there are few sellers in a particular region i.e, only a handful of farms, protecting location information becomes difficult as it can be easily linked to the corresponding farm location.\par 

The PrivChain design is based on an assumption that typical supply chains have broad geographical coverage. When there are a large number of producers in a region, it becomes harder to link transactions to individual producers and respective farm locations. However, linking attacks would still be possible in some cases and unavoidable.  
In such cases, regions with fewer farms can be merged with nearby regions with higher number of farms to operate under one side chain. Adjusting region boundaries can minimise the linking attacks, albeit at the cost of reduced resolution of region information and thus utility.

\subsubsection{Denial of Service Attacks}
In a permissioned blockchain network, only the registered participants can query/invoke transactions to launch denial of service attacks. For instance, a grape producer is able to launch a denial of service attack by sending a large number of $TX_{trade}$ or $TX_{create}$. \par
Supply chain perishable commodities such as grapes, are registered in a specific cycle, i.e., when the new produce is ready. Thus, the fake commodities can be readily detected if they are registered outside the expected cycle. In addition, a fake commodity cannot be traded, and hence can easily be identified on the blockchain. In case a large number of  $TX_{trade}$ are issued to gain more incentives, recall from Section~\ref{sec:arch} Step 3(a), that location proofs are verified once for each commodity at the time of trade and the validity status is stored in $TX_{trade}$. A blockchain peer does not allow ownership transfer or proof validation if the commodity has been traded in the past from the same seller. Thus, the commodity will only be traded once as a result of $TX_{trade}$ and the remaining transactions will not be successful. 



\subsubsection{Disputed Incentive Amounts}
During the incentive payment process as shown in Figure~\ref{fig:ver}, a grape producer and a wine producer agree on an incentive amount, $\$_{inc}$. However, a dishonest grape producer may choose a higher $\$_{inc}$ to generate $com_{inc}$. Similarly, a dishonest wine producer may choose a lower $\$_{inc}$ with an intention to pay less than the agreed amount. As a result, the off-chain verification at the bank will be invalid. The payments will be queued unless the disputes get resolved.\par
PrivChain does not allow payments if $\$_{inc} \neq \$'_{inc} $ at bank's end. Instead, a negative status of $Req_{pay}$ can be recorded on blockchain and payment transactions with a negative status can be traced back to $TX_{trade}$. If a seller or buyer is repeatedly involved in disputes related to incentives, their participation can be revoked.

\subsubsection{Replication of QR Code}
Recall from Section~\ref{sec:premsc}, after a wine bottle is registered on the blockchain, the printed provenance information is encoded in a QR code provided on the bottle. Since QR codes are easy to create and copy, counterfeiters may copy the original QR codes to use with a lower quality wine product.\par
The retailer information of the bottle is logged in the last $TX_{trade}$ before a bottle makes its way to the shelves. When an end consumer buys the bottle from the retailer, the status of the bottle gets updated automatically as ``sold" in the blockchain by the retailer. Thus, it is impossible for the retailer to sell bottles with the same QR code. 
In addition, the retailer does not have any incentive to sell fake bottles with copied QR codes as he can only sell one bottle associated with each QR code.
Alternatively, a Copy Detection Pattern (CDP)\citep{picard2004digital} along with QR codes can be employed to detect counterfeited wine bottles. A CDP is a maximum entropy image which exploits the property of information loss. A CDP image results in a loss of information when it is scanned and printed again from the original. A detector with an information loss threshold can determine the authenticity of a QR code with a CDP.

\subsubsection{Regional Blockchain Administrator}
Recall from Section~\ref{sec:arch}, a regional blockchain administrator devises most of the blockchain based tasks such as the access control and setup parameters for ZKRP. However, one may argue that excessive reliance on an administrative entity may pose an additional security risk to the system. A blockchain administrator may behave maliciously and jeopardise the security of the system.\par
Permissioned blockchains are particularly well suited for supply chains. Due to the nature of how permissioned blockchains work, the role of regional blockchain administrator cannot be evaded. In presence of audit and regulation it is not straightforward for administrative entity to behave maliciously. We argue that administrative rules must be decided by a consortium of entities such that no single entity can take control of the ledger. Consortium frameworks such as proposed in \citep{sjiang8946181, malik2018productchain}, allow multiple entities to devise and control the operations blockchain system.\par
In addition,the underlying cryptographic primitives of blockchain such as digital signatures, hash functions, and the data stored in transactions can prove the authenticity of information at the time of audit.
\section{Related Work}
Several data and identity privacy solutions have been proposed in the literature for decentralised applications.  Most of these solutions address privacy preservation by hiding the identity of the data provider on the blockchain while others focus on hiding the information shared. We discuss below certain solutions in literature which use encryption schemes, ZKPs, commitment schemes or multiparty computation to protect either data or identity over blockchains.
\subsection{Encryption Schemes}
 In \citep{pennekamp2020private}, the authors achieve privacy preservation by proposing distributed multi-hop accountability in which data can be accessed using Attribute-based Encryption (ABE). There are four main participants in the proposed system: access guards, collaborators, information storage providers and information coordinator. Collaborators are supply chain participants which store or retrieve information in a ledger. The access guards distribute ABE keys to the collaborators which then use ABE to encrypt the symmetric key used for encrypting raw data. The AES-encrypted raw data is kept at the storage provider and data link is provided later to the information coordinator. The information coordinator stores data submitted by a collaborator into a local database and the hash of this data is stored into the blockchain. Access guards can later distribute ABE attributes for authorised decryption of the raw data.  \par
In \citep{sjiang8946181}, authors have proposed  an efficient and privacy preserving bloom filter-enabled multi-keyword search protocol.
 The bloom filter selects a low-frequency keyword while performing a multi-keyword
search operation for encrypted data on blockchain. The low frequency key-word is selected for data filtration which limits the computational cost by returning small number of results. The authors have also proposed the use of pseudo random tags to further escalate the search process to be completed in a single round. The protocol outperforms the traditional search method in terms of time complexity and search cost.\par
 A recent work \citep{mitani2020traceability} on traceability proposes the use of HMM, homomorphic encryption and ZKPs for privacy preservation. The authors propose a permissioned blockchain connected to a permission-less blockchain. The privacy of information in the permissioned blockchain such as trade and financial transactions and the involvement of participants is not addressed. The proof of traceability requires calculations of more than quadratic degrees which are encrypted using homomorphic encryption. The number of participants in the permissioned blockchain corresponds to the number of additions made in the HMM. The model is encrypted using homomorphic encryption and the model establishment is later verified using ZKPs. \par

\subsection{ZKP and Commitment Schemes}

Another privacy preservation mechanism is proposed in \citep{li2018fppb} using stealth addresses and ZKP. The service providers register their services on the blockchain. The consumers after choosing the service, send a service request to the ledger masking their identity. A transaction verifies if a consumer has sufficient money to purchase the service. The verification is performed by dedicated participants. The service provider searches for the related request transactions, and commit response transactions. A consumer at some later stage can prove to a supervising authority the requests and money paid for the service. \par
A similar approach of hiding identities using ZKRP is adopted in \citep{gabay2020privacy} for electric vehicles. Electric vehicles need to schedule in-advance with a service provider as they take longer to charge. During the scheduling process, a service provider can learn private information such as location patterns, payment details, usage and habits. The proposed scheme issues scheduling and charging tokens to registered vehicles by respective smart contracts. These tokens are then used to avail the charging services at a charging station anonymously. The token scheme is, however, expensive in terms of blockchain transaction overhead \citep{gabay2020privacy}. The authors replaced the token based scheme with a Pedersen commitment scheme which showed a better performance.\par

In \citep{jiang2019privacy}, the authors use ZK-SNARK to prove that both the selling and buying parties have done payment and delivery of goods in escrow with a mediator, a registered regulator. Upon verification by the mediator, the funds are released to the seller and goods are transported using the delivery contract. 
The whole process is managed by three service oriented smart contracts: one for exchange of funds, second for adding updating the goods contract, and third to maintain delivery information of goods.\par

 Another identity anonymization approach is presented in \citep{uesugi2020short} where the blockchain addresses are encrypted along with the product distribution information on a public blockchain. The legitimate parties post manufacturing prove their identity using ZKP at the time of shipment and retrieval of product. Three contracts namely MMC, PMC and VC are used. MMC registers the manufacturer using its blockchain address and public key. MMC owner can then register its products with PMC, a contract which manages the ownership transfer requests. The owner first shares the secret token with the recipient, and using the same token, the owner encrypts the recipient address and manufacturer's public key. The encrypted recipient address and address of VC (which performs verification of identities) is then stored in PMC. The recipient generates a ZKP that it knows the secret token which was used to generate the encrypted address. PMC calls VC to verify if the recipient has a valid token and upon verification, it then grants the ownership of a product to the recipient. \par
 \subsection{MultiParty Computation}
 In \citep{benhamouda2019supporting}, authors argue that, since Hyperledger architecture aims to provide a consistent shared ledger among all the peers, it is challenging to keep information private. The authors propose a solution which uses Multi-Party Computation (MPC). The peers store encryption of private data on blockchain, and use MPC when a transaction requires this data. According to \citep{benhamouda2019supporting}, ZKP cannot be used in a setting where a smart contract depends on private data of more than one participant. In such cases, MPC is a better approach. In addition, the proof of concept was developed on Hyperledger for a bidding system where sellers have a secret reserve price for listed assets, and bidders publish their bids on the ledger by also keeping the bidding price secret. 
 
\begin{table*}[t] 
\caption{Comparison of privacy preservation approaches in blockchain enabled applications}
  \label{tab:comp}
\begin{center}
\footnotesize
\begin{tabular}{llcccccc}

\hline
\textbf{Article Name} & \textbf{Use-Case}                                                                   & \multicolumn{1}{c}{\textbf{Scheme}}                                  & \multicolumn{1}{c}{\textbf{Platform}}                                  & \multicolumn{1}{l}{\textbf{\begin{tabular}[c]{@{}l@{}}Identity\\  protection\end{tabular}}} & \multicolumn{1}{l}{\textbf{\begin{tabular}[c]{@{}l@{}}Data \\ protection\end{tabular}}} & \multicolumn{1}{l}{\textbf{Incentive}} & \multicolumn{1}{l}{\textbf{\begin{tabular}[c]{@{}l@{}}Privacy \\ Analysis\end{tabular}}} \\ \hline
 Pennekamp et al.\citep{pennekamp2020private}      & \begin{tabular}[c]{@{}l@{}}Supply Chains\\ (traceability)\end{tabular}              & ABE, CP-ABE                                                          & \begin{tabular}[c]{@{}c@{}}Permissioned\\ (Quorum)\end{tabular}        & N                                                                                           & Y                                                                                       & N                                      & Y                                                                                        \\
Li et al. \citep{li2018fppb}                  & Cloud Computing                                                                     & Stealth addresses, PC                                                & \begin{tabular}[c]{@{}c@{}}Permissioned\\ (Ethereum J)\end{tabular}    & Y (partial)                                                                                 & N                                                                                       & N                                      & Y                                                                                        \\
Gabay et al.\citep{gabay2020privacy}         & \begin{tabular}[c]{@{}l@{}}Electric Vehicle \\ (authentication )\end{tabular}     & PC, ZKP                                                              & \begin{tabular}[c]{@{}c@{}}Public\\ (Ethereum)\end{tabular}            & Y                                                                                           & N                                                                                       & N                                      & Y                                                                                        \\
Jiang et al. \citep{jiang2019privacy}         & \begin{tabular}[c]{@{}l@{}}Supply Chain\\ (trading)\end{tabular}                    & ZKP                                                                  & \begin{tabular}[c]{@{}c@{}}Permissioned\\ (Quorum)\end{tabular}        & Y                                                                                           & Y(partial)                                                                              & N                                      & N                                                                                        \\
Benhamouda et al. \citep{benhamouda2019supporting}            & \begin{tabular}[c]{@{}l@{}}Supply Chain\\ (auctioning)\end{tabular}                 & MPC                                                                  & \begin{tabular}[c]{@{}c@{}}Permissioned\\ (Hyperledger)\end{tabular}   & N                                                                                           & Y                                                                                       & N                                      & Y                                                                                        \\
S. Jiang et al \citep{sjiang8946181}          & \begin{tabular}[c]{@{}l@{}}Supply Chain\\ (database search)\end{tabular}               & \begin{tabular}[c]{@{}c@{}}Bloom Filters, \\ symmetric encryption] \end{tabular} & \begin{tabular}[c]{@{}c@{}} encrypted database\end{tabular} & N                                                                                           & Y                                                                             & N                                      & N                                                                                     \\

A. Mitani \citep{mitani2020traceability}          & \begin{tabular}[c]{@{}l@{}}Supply Chain\\ (traceability)\end{tabular}               & \begin{tabular}[c]{@{}c@{}}HMM, ZKP\\ Hommomorphic Enc.\end{tabular} & \begin{tabular}[c]{@{}c@{}}Permissioned,\\ Permissionless\end{tabular} & Y                                                                                           & Y(partial)                                                                              & N                                      & Y                                                                                        \\
PrivChain             & \begin{tabular}[c]{@{}l@{}}Supply Chains\\ (provenance\\ traceability)\end{tabular} & ZKP, PC, AES                                                         & \begin{tabular}[c]{@{}c@{}}Permissioned \\ (Hyperledger)\end{tabular}  & Y(partial)                                                                                           & Y                                                                                       & Y                                      & Y                                                                                        \\ \hline
\end{tabular}
\end{center}

\end{table*}

\subsection{Summary and Challenges}
In summary, the solutions discussed above address some privacy challenges in context of blockchains but most of them do not address the problem of privacy while simultaneously achieving traceability and provenance. The work closest to the objectives of PrivChain is presented in \citep{pennekamp2020private} and \citep{mitani2020traceability}. The novelty of PrivChain stems from its suitability for data protection, provenance and traceability provision without the need of masking identities, information managing peers or need for decryption to retrieve provenance information. Some significant features of PrivChain compared to the work proposed in \citep{pennekamp2020private,mitani2020traceability} are: a) the data always stays with the data owner; b) the verification of provenance data is pre-computed before a final product is registered, thus minimising the query time for provenance; c) un-linkability of the trade flows involved in a final product;  d) incentive mechanism for data owners on providing valid proofs; and e) introduction of minimal overheads to the baseline system. In addition, the design of PrivChain can be generalised for any type of data and supply chain use-case. Moreover, Table \ref{tab:comp} provides a high level comparison of  PrivChain with the existing literature discussed in this section. The incentive mechanism is clearly a unique feature of PrivChain compared to other approaches, which ensures that participants are motivated to incur the computational cost of providing verified proofs.

\section{Discussions}
In the following sections, we discuss the complexity of using ZKRPs in blockchain enabled systems and the general application of PrivChain for other use-cases apart from supply chain.
\subsection{Complexity of Using ZKRPs for blockchain enabled applications}
The computational complexity of the ZKRPs is discussed in Section~\ref{sec:premzkp}. Moreover, the time complexity is evaluated in Figure~\ref{fig:zkrpOH} which shows that proof verification overhead of ZKRPs in blockchain, is low. This is because the data used in ZKRP is numeric and the consensus mechanism of Hyperledger Fabric (RAFT, Kafka) is not as computationally expensive as Proof of Work. Thus any consortium based blockchain application can cater to the needs of resources used for setup (off-chain) and proof verification (on-chain). In addition, given the proof generation time, resource constrained devices can perform these computations. However, the willingness for resource consumption for proof generation is relative to the application scenario. For applications where the resources may be limited for the end-users, PrivChain incorporates an incentive mechanism for reciprocation of resource utilization at the user's end.\par

In general, the computational complexity of ZKPs depends on the underlying variant of zero knowledge proof, data to be used in generation of proofs and the resources available to end users, which is contingent on the specific application.

\subsection{Generality of PrivChain} 
The proposed PrivChain framework is generic and can be used to prove any numeric range data stored on a permissioned blockchain without revealing the data itself. For example, it can be used to prove if the account balances lie within a certain threshold while making a payment. Alternately, zero knowledge set membership protocols can be used to prove other non numeric data. For example credential verification is another application of ZKP which requires a user to prove their credentials to the requester. Such a credential proving system is realised by Hyperledger Indy. Recall from Section~\ref{sec:intro}, that PrivChain framework can be applied in any decentralised application where provenance and privacy cannot be acheived in parallel. The following use case demonstrates PrivChain application in energy trading.

\begin{figure}[t]
  \centering
  \includegraphics[width=7.5cm]{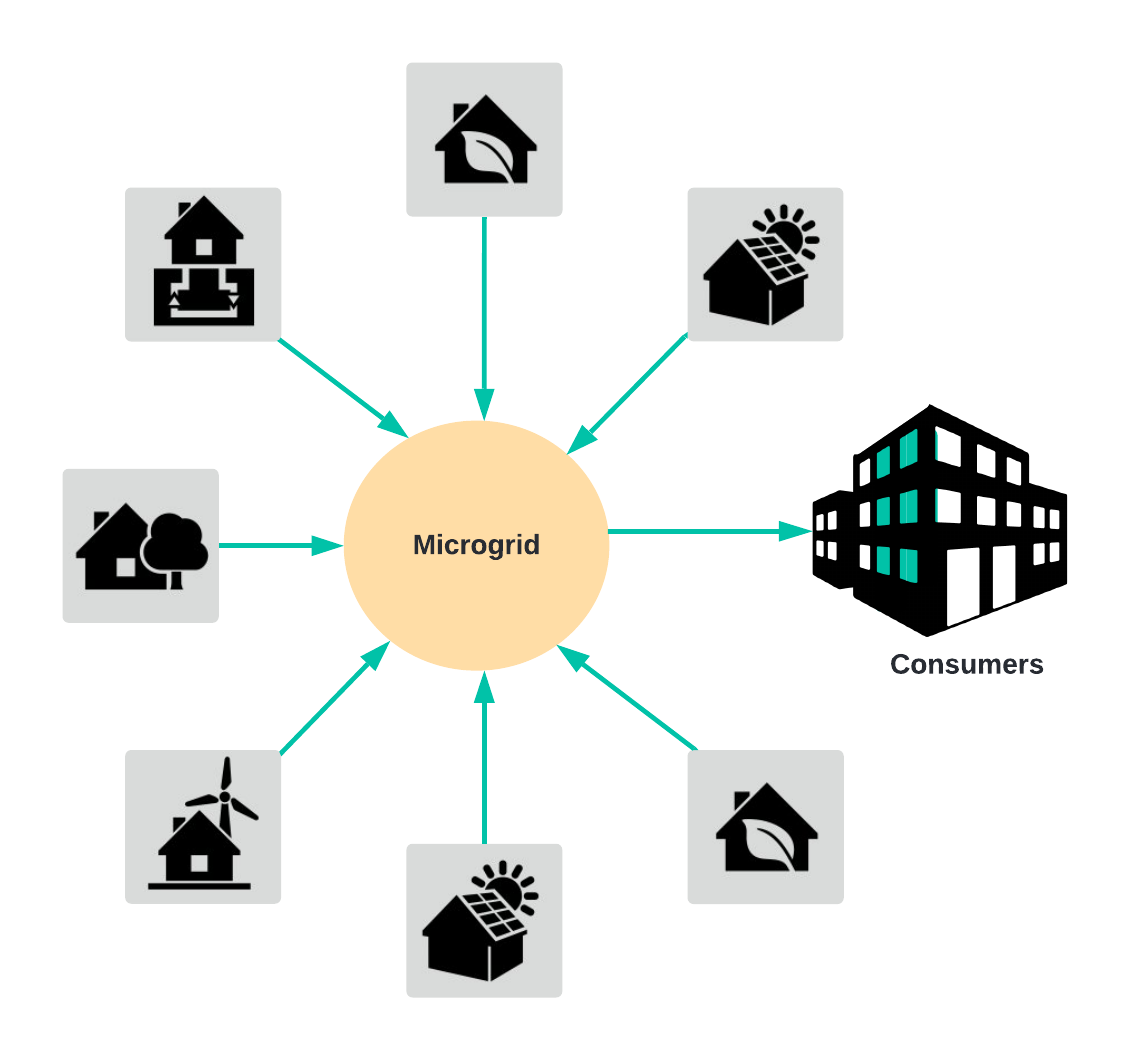}
  \caption{A Use-case of Energy Trading}
  \label{fig:genergy}
\end{figure}

 With the advent of IoT and Distributed Energy
Resources (DERs), (e.g solar PV, battery, electric vehicles etc) energy production and consumption load is widely managed through Peer to Peer Decentralised Energy Trading (P2P DET)\citep{karumba2020harb,privacy_smartgrid}. This has given rise to the role of prosumers, i.e. a producer-and-consumer of energy for example, a residential smarthome equipped with a solar panel.  Blockchain is adopted as a promising solution in P2P DET due to its salient features such as immutability which could support provenance and traceability of DET data. To explain PrivChain application in DET, consider a simple use-case of green energy production by prosumers of smart home.  Smart homes are registered producers of this green energy using different resources i.e. solar, wind, bio-gas, stored energy etc. as shown in Figure~\ref{fig:genergy}. The green energy is then sold to the micro-grid which further sells it to the energy consumers. Consumers may be interested to know if the source of energy was from renewable resources and thus require a proof of provenance. In this particular use-case, blockchain is an ideal solution to prove the provenance of the energy resources. However, the prosumers will likely want to hide their energy production patterns to preserve their privacy, while proving that the energy provided to the micro-grid is green. The PrivChain framework can be applied here to hide either the location of smart homes or the type of renewable energy sources or both using ZKPs. The micro-grid can verify the proofs and the verification results can be stored on the blockchain for consumer verification. Following a valid proof verification, the payments plus proof generation incentives can be provided by the micro-grid through tokens which could be redeemed from a third party financial institution.

\section{Conclusion}
In this paper, we proposed a privacy preservation framework, PrivChain,
for blockchain enabled supply chain applications. PrivChain aims to protect the trade secrets of supply chain entities such as the data related to location and trade flows, without losing the the provenance and traceability features. To address the data privacy problem, PrivChain uses zero knowledge proofs, where supply chain entities can share proofs instead of the actual data and are incentivised for providing valid proofs. In particular, we applied the PrivChain design on a use-case of wine supply chains. The grape producers provide location proofs using ZKRPs which ascertain that the grapes are sourced from a a particular wine-growing region without revealing the precise location. PrivChain framework also ensures the grape producers are paid agreed incentive amounts by the wine producers for providing proofs. In addition, to protect the trade flows involved in the production of a wine bottle, the respective ingredient identifiers are encrypted by the wine producers. The implemented framework in Hyperledger incurs minimal overheads for proof verification, and encryption related tasks. We also performed a qualitative security analysis with respect to some known threats applicable to the PrivChain framework. PrivChain framework can be adapted to use ZKRP for proving many numeric calculations over decentralised applications without revealing actual data, such as account balances, number of sales, sensor readings, ratings, etc. For future work, we aim to explore other variants of zero knowledge proofs such as Bulletproofs.

\bibliographystyle{abbrv}
\bibliography{main.bib}

\vspace{2cm}
\noindent \textbf{Sidra Malik} is a Ph.D. student at School of Computer Science and Engineering, UNSW and  Distributed Sensing Systems Group of CSIRO Data61, Australia. Her research work focuses on providing solutions regarding traceability, trust and privacy for blockchain-enabled supply chains .
	Before commencing her Ph.D., Sidra worked as a Lecturer at COMSATS University, Pakistan. She has also been working as Software Quality Assurance Engineer at Techlogix, Pakistan. She received her MS degree in Computer and Communication Security from National University of Science and Technology, Pakistan (2012).
	Her BS is in Computer and Information Sciences from Pakistan Institute of Engineering and Applied Sciences (2009).
\subsection*{  } 
\noindent \textbf{Volkan Dedeoglu}  is currently a postdoctoral research fellow in the Distributed Sensing Systems Group of CSIRO Data61. His current research focuses on data trust and blockchain-based IoT security and privacy. Volkan also holds Adjunct Lecturer positions at UNSW Sydney and QUT. Before joining Data61, Volkan worked as a postdoctoral research associate at Texas A\&M University on physical layer security for communications. He completed his PhD in Telecommunications Engineering from UniSA (Australia, 2013) on energy efficient wireless sensor networks, data gathering and target tracking. He obtained MSc in Electrical and Computer Engineering from Koc University (Turkey, 2008), BSc in Electrical and Electronics Engineering from Bogazici University (Turkey, 2006). 
\subsection*{  } 
\noindent \textbf{Salil S Kanhere}
received his M.S. degree in 2001 and Ph.D. degree in 2003 from Drexel University, Philadelphia. He is a Professor of Computer Science and Engineering at UNSW Sydney, Australia. His research interests include Internet of Things, blockchain, pervasive computing, cybersecurity and applied machine learning. He is a Senior Member of the IEEE and ACM and an ACM Distinguished Speaker. He is a recipient of the Friedrich Wilhelm Bessel Research Award (2020) and the Humboldt Research Fellowship (2014), both from the Alexander von Humboldt Foundation. He serves as the Editor in Chief of the Ad Hoc Networks journal and as Associate Editor of IEEE Transactions on Network and Service Management, Computer Communications and Pervasive and Mobile Computing. He has served on the organising committee of many IEEE/ACM international conferences including PerCom, CPS-IOT Week, MobiSys, WoWMoM, MSWiM, ICBC.

\subsection*{  } 
\noindent \textbf{Raja Jurdak}
	is a Professor of Distributed Systems and Chair in Applied Data Sciences at Queensland University of Technology, and Director of the Trusted Networks Lab. He received the PhD in information and computer science from the University of California, Irvine in 2005. He previously established and led the Distributed Sensing Systems Group at CSIRO’s Data61, where he maintains a visiting scientist role. His research interests include trust, mobility and energy-efficiency in networks. Prof. Jurdak has published over 180 peer-reviewed publications, including two authored books most recently on blockchain in cyberphysical systems in 2020. He serves on the editorial board of Ad Hoc Networks, and on the organising and technical program committees of top international conferences, including Percom, ICBC, IPSN, WoWMoM, and ICDCS. He is a conjoint professor with the University of New South Wales, and a senior member of the IEEE.

\end{document}